\documentclass[prb,twocolumn,superscriptaddress,shownopacs]{revtex4}

\usepackage{graphicx}
\usepackage{hyperref}
\usepackage{amsmath}
\usepackage{epstopdf}
\usepackage{bbm}
\newcommand{\ket}[1]{\vert #1 \rangle}
\usepackage{xcolor}
\usepackage{siunitx}
\usepackage{ulem}

\begin{document}

\title{Storage and retrieval of vector beams of light\\in a multiple-degree-of-freedom quantum memory}

\author{Valentina Parigi}
\affiliation{Laboratoire Kastler Brossel, UPMC-Sorbonne Universit\'es, CNRS, ENS-PSL Research University, Coll\`ege de France, 4 place
Jussieu, 75005 Paris, France}
\author{Vincenzo D'Ambrosio\footnotemark[2]\footnotetext{\footnotemark[2]These authors contributed equally to the work.}}
\affiliation{Dipartimento di Fisica, ``Sapienza''
Universit\`{a} di Roma, I-00185 Roma, Italy}
\author{Christophe Arnold\footnotemark[2]}
\affiliation{Laboratoire Kastler Brossel, UPMC-Sorbonne Universit\'es, CNRS, ENS-PSL Research University, Coll\`ege de France, 4 place
Jussieu, 75005 Paris, France}
\author{Lorenzo Marrucci}
\affiliation{Dipartimento di Fisica, Universit\`a di Napoli Federico II,
Complesso Universitario di Monte S. Angelo, 80126 Napoli, Italy}
\affiliation{CNR-SPIN, Complesso Universitario di Monte S. Angelo, 80126 Napoli, Italy}
\author{Fabio Sciarrino}
\affiliation{Dipartimento di Fisica, ``Sapienza''
Universit\`{a} di Roma, I-00185 Roma, Italy}
\author{Julien Laurat}
\email{julien.laurat@upmc.fr}
\affiliation{Laboratoire Kastler Brossel, UPMC-Sorbonne Universit\'es, CNRS, ENS-PSL Research University, Coll\`ege de France, 4 place
Jussieu, 75005 Paris, France}

\date{\today} 
\maketitle
 
\noindent \textbf{The full structuration of light in the transverse plane, including intensity, phase and polarization, holds the promise of unprecedented capabilities for applications in classical optics as well as in quantum optics and information sciences. Harnessing special topologies can lead to enhanced focusing, data multiplexing or advanced sensing and metrology. Here we experimentally demonstrate the storage of such spatio-polarization-patterned beams into an optical memory. A set of vectorial vortex modes is generated via liquid crystal cell with topological charge in the optic axis distribution, and preservation of the phase and polarization singularities is demonstrated after retrieval, at the single-photon level. The realized multiple-degree-of-freedom memory can find applications in classical data processing but also in quantum network scenarios where structured states have been shown to provide promising attributes, such as rotational invariance.}

Vector beams of light constitute the class of beams characterized by a space-variant polarization in the transverse plane \cite{maurer2007tailoring}. Among them, an important subclass are those having cylindrical-symmetric polarization patterns, including radial, azimuthal and spiraling polarizations \cite{zhan2009cylindrical,Fick14}. These beams can be expressed as combinations of two twisted waves, i.e. doughnut-shaped Laguerre-Gaussian modes, with opposite orbital-angular-momentum (OAM) topological charge $l=+1$ and $l=-1$ and opposite uniform circular polarizations. Combining polarization and coincident phase singularities, these states are sometimes called vector vortex beams \cite{Cardano2012}. Symmetric pairs of vector beams define two-dimensional spaces of non-uniform polarization states formally analogous to the standard Poincar\'e sphere, which are known as ``hybrid Poincar\'e spheres'' or ``higher-order Poincar\'e spheres'' \cite{Souz07,Holl11,milione2011higher}. 

Over the recent years, vector beams have raised a large interest as spatially arranging the polarization profile opens the possibility to tailor the magnetic and electric field distribution in their focus. Applications cover a wide range of areas, including particle acceleration \cite{kimura1995laser} and trapping \cite{roxworthy2010optical}, sharper focusing \cite{dorn2003sharper,Leuchs2014}, improved laser cutting and drilling \cite{nesterov2000}, enhanced microscopy \cite{abouraddy2006three} and metrology \cite{DAmbrosio2013}. 

Besides their applications in classical optics, vector beams unusual attributes have also been exploited to investigate quantum mechanics foundations and their use as a novel resource in quantum information protocols triggered a variety of seminal demonstrations. Indeed such beams correspond to states living in a high-dimensional hybrid Hilbert space based on polarization and orbital angular momentum, allowing to encode single-photon qudits. For instance, these states have been used to test the Hardy's paradox \cite{Kari14} and to investigate experimentally the violation of inequalities based on quantum contextuality \cite{d2013experimental}. In quantum information, vector beams have already found a large range of applications, such as performing quantum walks and simulations \cite{Card14}, generating novel kind of cluster states  \cite{Gabriel2011,Rigas2012} or implementing and testing mutually unbiased bases in high-dimensional spaces \cite{d2013test}. Moreover, a particular class of vector beams exhibits the interesting property of being rotationally-invariant and has been recently exploited for alignment-free quantum communication between distant parties \cite{DAmbrosio2012,vallone2014free,Aoli07}. This technique allows to overcome communication errors due to reference frame misalignments that can affect the security of cryptograpic protocols.

Here we report on the storage and retrieval of vector beams at the single-photon level, using a multiplexed ensemble of laser-cooled atoms. This setting requires the simultaneous reversible mapping of the polarization and spatial degrees of freedom, a capability which has not been addressed until now. In our experiment, a set of vortex beams are generated with a patterned birefringent liquid crystal plate and mapped into and out of the memory cells via a dynamically controlled protocol \cite{Harris,Fleischhauer,Lvovsky2009}.  Moreover, we demonstrate the preservation of the rotational invariance of the patterned beams, with fidelities close to unity and exceeding classical benchmarks for memory protocols. Our work thereby provides a quantum register for hybrid polarization-OAM states, suitable for quantum information protocols based on vector beams, including alignement-free quantum communications \cite{DAmbrosio2012,vallone2014free}.

\begin{figure}[t!]
\begin{center}
\includegraphics[width=0.9\columnwidth]{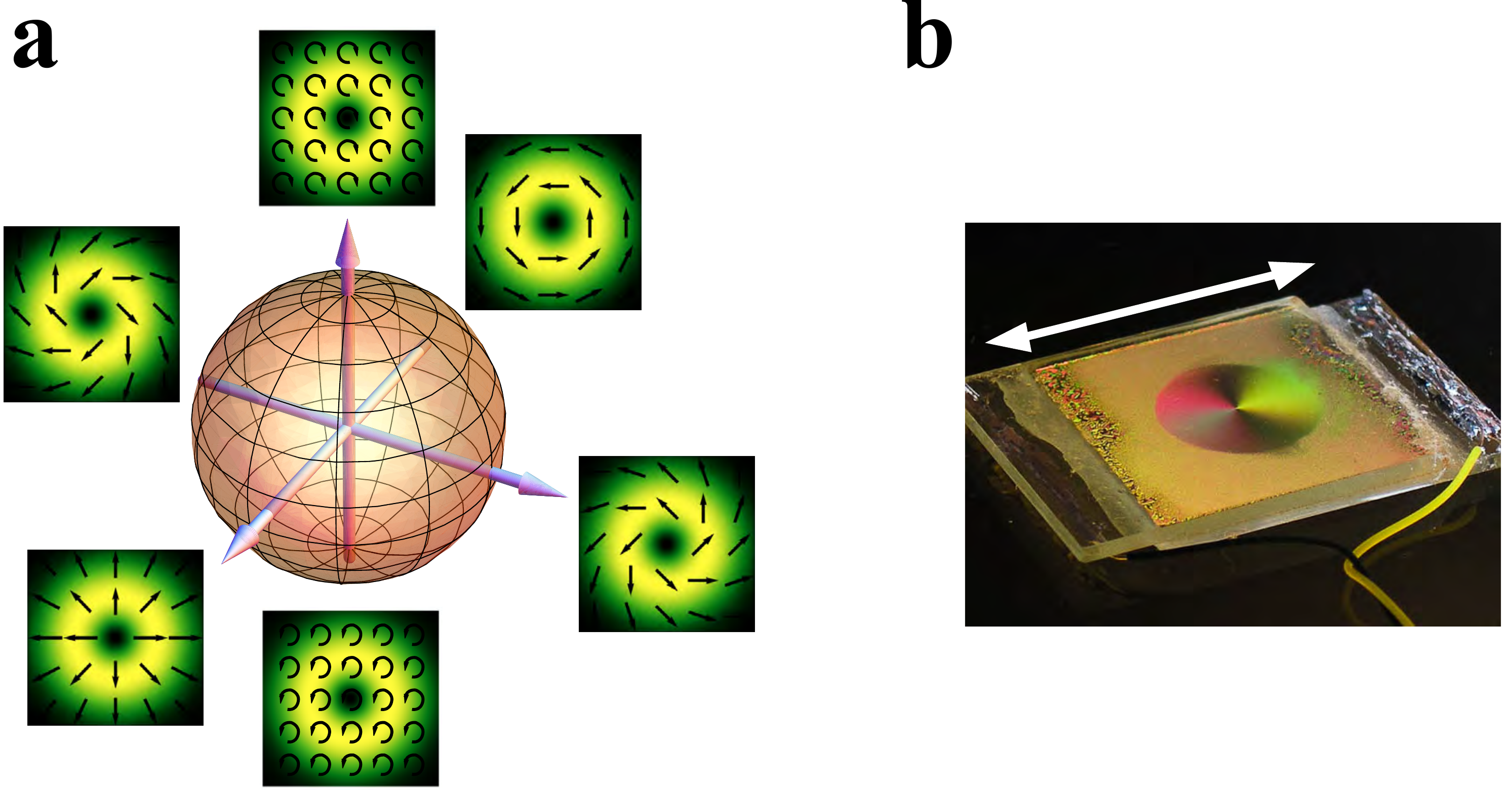}
\end{center}
\caption{\textbf{Vector beam representation and generation.} (a) Vector beam qubit can be represented in a hybrid Poincar\'e sphere, where the south and north poles are respectively the states $\ket{0}=\ket{L,-1}$ and $\ket{1}=\ket{R,+1}$ with opposite circular polarizations and orbital angular momenta. Linear combinations of these logical states give rise to special polarization topologies in the transverse plane, including radial, azimutal and spiraling polarization patterns. (b) Picture of a $q$-plate with topological charge $q=1/2$ at its center, seen through crossed polarizers and under oblique illumination. The transverse size is about 2 cm. This patterned liquid crystal cell enables to couple spin and orbital angular momentum. Any polarization qubit can be mapped into cylindrical vector beams and reciprocally. The retardation is fine-tuned by external voltage.} \label{vectorbeams}
\end{figure}

Cylindrical-symmetric polarized beams can be expressed as linear combinations of two polarization-OAM hybrid states, $\ket{0}=\ket{L,-1}$ and $\ket{1}=\ket{R,+1}$, where $R$ and $L$ represent the opposite right and left circular polarizations and the second term indicates the orbital angular momentum of the photon state in $\hbar$ units. These two states constitute the north and south poles of the hybrid Poincar\'e sphere, as shown on Fig. \ref{vectorbeams}a. For instance, radially and azimuthally polarized photons lie in the equatorial plane, corresponding to in-phase or out-of phase combination. These basis states, and any superpositions, can be easily generated by exploiting a liquid-crystal-based device, called $q$-plate as pictured in Fig. \ref{vectorbeams}b, which is able to couple spin and orbital angular momentum \cite{Marrucci2006,Piccirillo2010,Marrucci2011}.

More specifically, a $q$-plate is a birefringent plate whose optical axis transverse distribution shows a singular pattern with topological charge $q$. In polar coordinates the optic axis orientation is described by $\alpha(r,\varphi) = q \varphi + \alpha_0$ where the plate lies on the $\textit{xy}$ plane, $\alpha$ is the angle formed by the optic axis with the $\textit{x}$ axis and $\alpha_0$ is a constant offset angle. Given an impinging qubit in the generic polarization state $\ket{\psi}=\alpha\ket{R}+\beta\ket{L}$, a $q$-plate with topological charge $q$ maps it into:
\begin{equation}
\ket{\psi}\longrightarrow\alpha\ket{L,-2q}+\beta\ket{R,2q}.
\end{equation}
Hence radial and azimuthal vector beams can be easily obtained by injecting respectively a linear horizontal and vertical polarized beam in a $q$-plate with charge $q=0.5$, as used here. All the states on the corresponding hybrid Poincar\'e sphere can be generated by simply acting on the polarization of the input beam. On the other hand, a vector state can be converted to a polarization state and analyzed by exploiting again a $q$-plate and standard polarization optics:
\begin{equation}
\alpha\ket{L,-2q}+\beta\ket{R,2q} \longrightarrow \alpha\ket{R}+\beta\ket{L}.
\end{equation}
Thereby, the manipulation and analysis of vector beams can be completely performed in the polarization space, the $q$-plate being an interface between polarization space and cylindrical vector beams space\cite{DAmbrosio2012,DAmbrosio2013}. 

In the present work, the vector beams are implemented at the single-photon level, using attenuated coherent states with a mean-photon number per pulse equal to $\overline{n}=0.5$.

\begin{figure*}[t!]
\begin{center}
\includegraphics[width=1.7\columnwidth]{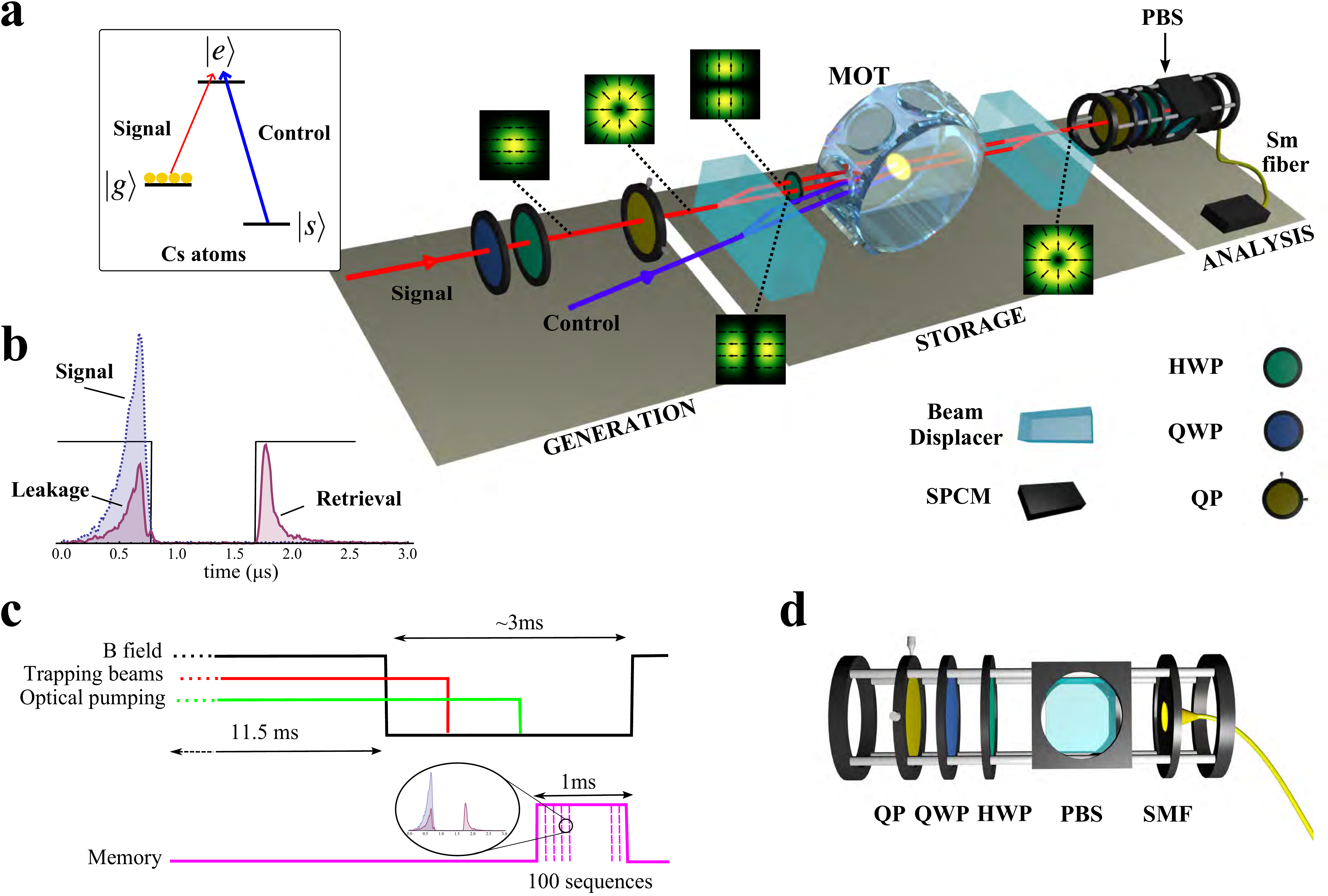}
\end{center}
\caption{\textbf{Multiple-degree-of-freedom quantum memory for vector beam storage.} (a) Vector beams, at the single-photon level, are first generated from polarization qubit via $q$-plate (QP) and then coherently mapped into and out of a dual-rail multiplexed quantum memory based on a single large ensemble of laser-cooled cesium atoms. The two polarization paths are defined by calcite beam displacers and separated by 650 $\mu$m. The Gaussian control beam and the vector beams to be stored copropagate with an angle of 3$^{\circ}$. After on-demand retrieval, the state is converted back to the polarization space thanks to a second $q$-plate and usual polarization tomography is then performed. As an illustration, intensity and polarization profiles for an impinging Gaussian H-polarized beam is shown: polarization projection by the beam-displacer results into two Hermite-Gaussian modes, rotated by 90$^{\circ}$ and orthogonally polarized. The left inset shows the involved energy levels. (b) Typical memory experiment, with input light in the absence of atoms in blue, leakage and retrieval in red, and control timing in black. (c) Timing of the experiment. 100 sequences are repeated at each MOT cycle. (d) Details of the analysis cage enabling decoding and subsequent polarization tomography. MOT, magneto-optical trap; PBS, polarizing beam splitter; SMF, single mode fiber; HWP, half wave plate; QWP, quarter wave plate; SPCM, single-photon counting module.} \label{setup}
\end{figure*}

Storing complex vector states requires the capability to reversibly map multiple degrees of freedom, i.e. polarization and spatial mode, while preserving all the coherences involved. The experimental setup is illustrated on Fig. \ref{setup}a. Vector beams are stored in a large ensemble of cold cesium atoms prepared inside a magneto-optical trap (see Methods). The relevant atomic $\Lambda$-system for the reversible mapping is based on two hyperfine ground states $\ket{g}=\ket{6S_{1/2},F=4}$ and $\ket{s}=\ket{6S_{1/2},F=3}$, and one excited state $\ket{e}=\ket{6P_{3/2},F=4}$. Cesium atoms are first prepared in the $\ket{g}$ state and no additional pumping in a specific Zeeman sub-level is performed. We have spectroscopically checked that all the magnetic sub-levels of the F=4 manifold are significantly populated.  Residual magnetic fields are compensated down to 5 mG in order to attain the frequency degeneracy of all Zeeman sub-levels (see Methods). The light to be stored adresses the $\ket{g}\leftrightarrow\ket{e}$ transition. 

A control beam on the $\ket{s} \leftrightarrow \ket{e}$ transition is first shined on the atomic ensemble, with a waist of 400 $\mu$m. By switching off this beam, the optical states are then coherently mapped onto a collective spin excitation. Switching on the control at a later time allows the retrieval of the signal in the same spatial mode thanks to a collective enhancement effect. Signal pulses, with a beam waist equal to 50 $\mu$m, are temporally shaped with an exponential rising profile, with a full-width at half maximum equal to 200 ns, close to the time-reversal of the retrieved pulses\cite{Novikova2007} (Fig. \ref{setup}b). The memory efficiency is optimized by detuning the control by 10 MHz from resonance. This detuning enables to mitigate the residual absorption of the re-emitted signal by the dense atomic cloud.  We work thus in an intermediate configuration between EIT \cite{Fleischhauer, Lvovsky2009} and off-resonant Raman schemes \cite{Ian}. This configuration requires the adjustment of the two-photon detuning in order to maximize the memory efficiency and we experimentally found an optimal value equal to 100 kHz.

\begin{figure*}[t!]
\begin{center}
\includegraphics[width=1.7\columnwidth]{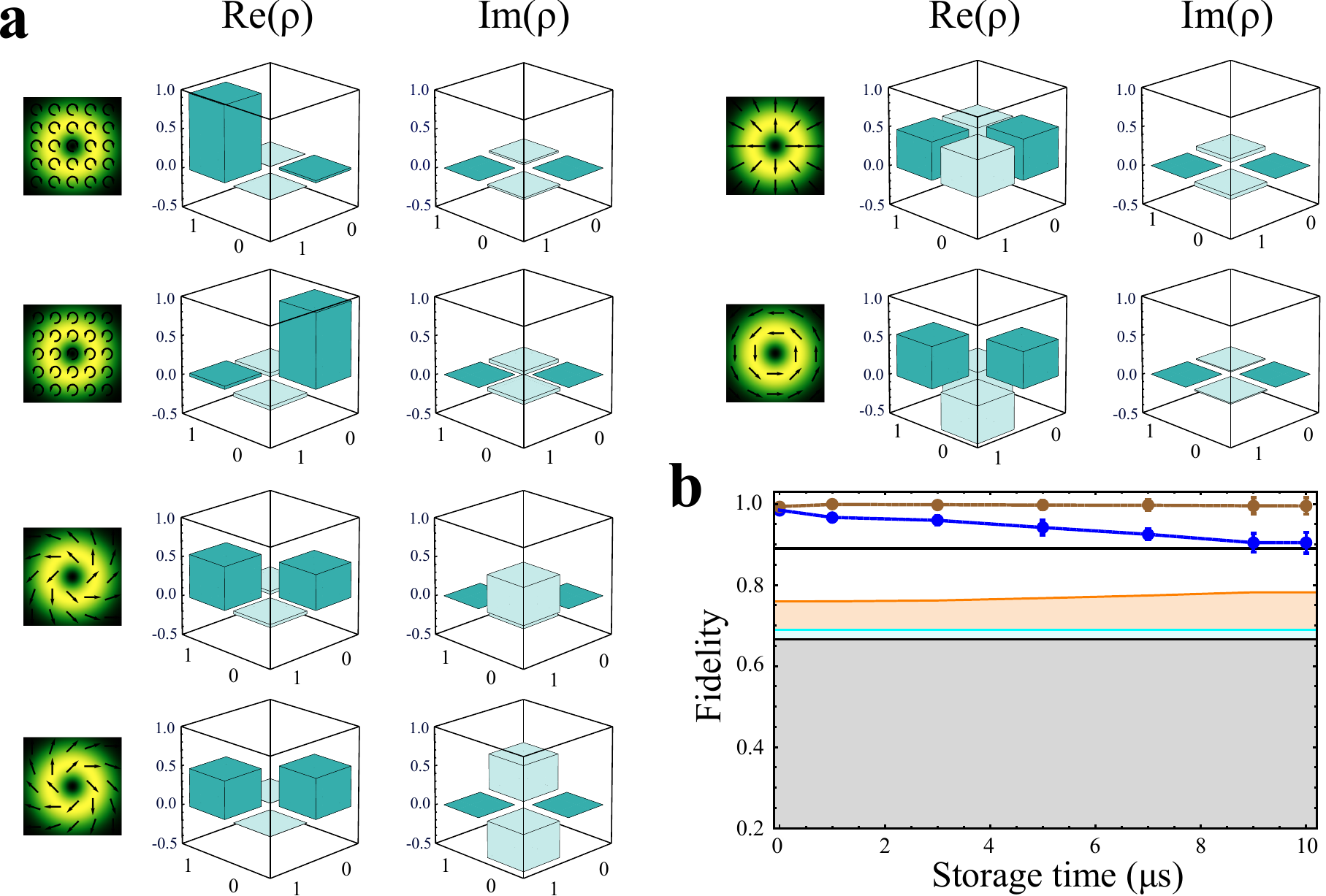}
\caption{\textbf{Quantum tomography of the retrieved vector beams of light.} (a) Real and imaginary parts of the density matrix of the retrieved states plotted in the \{$\vert 0\rangle$,$\vert 1\rangle$\} basis after a 1 $\mu$s storage time. (b) Average fidelity over the six retrieved states as a function of the storage time. Blue points correspond to raw data while the the brown points are corrected for background noise. Without correction, the fidelity stays above the Shor-Preskill threshold ($0.89$, black solid line) as long as the signal to noise ratio keeps a value higher than $8$. The gray area indicates the fidelity values achievable with a classical memory protocol for a single photon, limited thus to $2/3$, the light blue extension takes into account the Poissonian statistics of the weak coherent state and the light orange one finally includes the finite retrieval efficiency (see Methods). The mean number of photons per pulse is here $\overline{n}=0.5$. The vertical error bars indicate the standard deviation of fidelities for the six stored states.}\label{tom}
\end{center}\end{figure*}

The intrinsic multimode character of ensemble-based memory implementation enables to preserve the spatial phase and intensity distribution, as recently demonstrated for twisted light \cite{Veissier2013,Ding2013,Nicolas2014}. However, spatial polarization variation of the vector beams requires polarization independence for the memory protocol, which is not the case here as control and signal need to be orthogonally polarized in our atomic configuration. This stringent requirement can however be achieved by using a dual-rail memory strategy. Accordingly, signal and control beams go through two beam-displacers based on birefringent calcite crystals in order to achieve polarization multiplexing \cite{Kuzmich04, Chou07, Laurat07}: the first one is placed before the memory and separates each of them into two beams with well-defined H and V polarizations. A half-wave plate acts only on the signal path after the first displacer. With this method, we ensure that each polarization projection of the signal beam is superimposed with a control beam with the proper polarization. The two paths are separated by 650 $\mu$m, which is smaller than the transverse size of the sample and allows to get a balanced optical depth of 12 on each path, leading to an overall storage and retrieval efficiency of $\eta=(26\pm1)\%$, as shown in Fig. \ref{setup}b. The efficiency is defined as the ratio of the photodetection event probability in the read-out to the one in the reference, i.e. without atoms loaded in the trap. The second displacer finally recombines the two polarization components of the retrieved vector beam. The relative phase between the two interferometric paths is set to zero by adjusting the tilt of one of the beam displacers and no active phase stabilization is required during the experiment as this configuration can be stable for hours.

The capability of storing vector states is first proved by performing quantum state tomography of the states retrieved after storage into the atomic memory. The two states $\vert 0\rangle$ and $\vert 1\rangle$, i.e. the two poles of the hybrid Poincar\'e sphere, and four states on the equator are generated using the encoding stage, which includes birefringent waveplates and a $q$-plate. As an illustration, intensity profile and polarization of the signal at different positions in the experiment is shown on Fig. \ref{setup}a for a H-polarized gaussian beam impinging on the $q$-plate. These states are then retrieved from the memory and reconverted to polarization states by the second q-plate in the analysis path. By operating six independent projective measurements by the usual combination of waveplates, polarizer and single-photon counter module, the Stokes parameters are experimentally evaluated and the density matrix of the state is reconstructed.

The results are shown in Fig. \ref{tom}a, where the real and imaginary part of the reconstructed density matrix are plotted in the \{$\vert 0\rangle$,$\vert 1\rangle$\} logical basis. The conditional fidelity calculated between the retrieved states and the target state is calculated according to $F_c=\langle \psi\vert \hat{\rho} \vert \psi \rangle $, where $\hat{\rho}$ is the measured state and $\vert \psi \rangle $ the ideally encoded state. The  average value over the six input states is $F_c=(96.7 \pm 0.7)\%$  if the tomography is performed using raw data, and it reaches $F_c=(99.5 \pm 0.5)\%$  if the residual background noise coming from dark counts and residual control leakage is subtracted. This value corresponds to the upper bound of the maximal attainable fidelity, given by fidelity of the encoding-decoding process $F_{ed}=(99 \pm 1)\%$ measured with the six hybrid states in the case of intense beams and without atoms. The fidelity, which benefits from the long-term phase stability of the Mach-Zehnder interferometer formed by the two displacers, is not reduced by the atomic storage, preserving the complex pattern of the vector beam. As shown in Fig. \ref{tom}b, the demonstrated fidelities are above the limits given by the best achievable classical memory protocol (see Methods), known for instance in cryptography as intercept-resend scenario, demonstrating therefore the quantum character of our memory implementation.

\begin{figure*}[t!]
\begin{center}
\includegraphics[width=1.8\columnwidth]{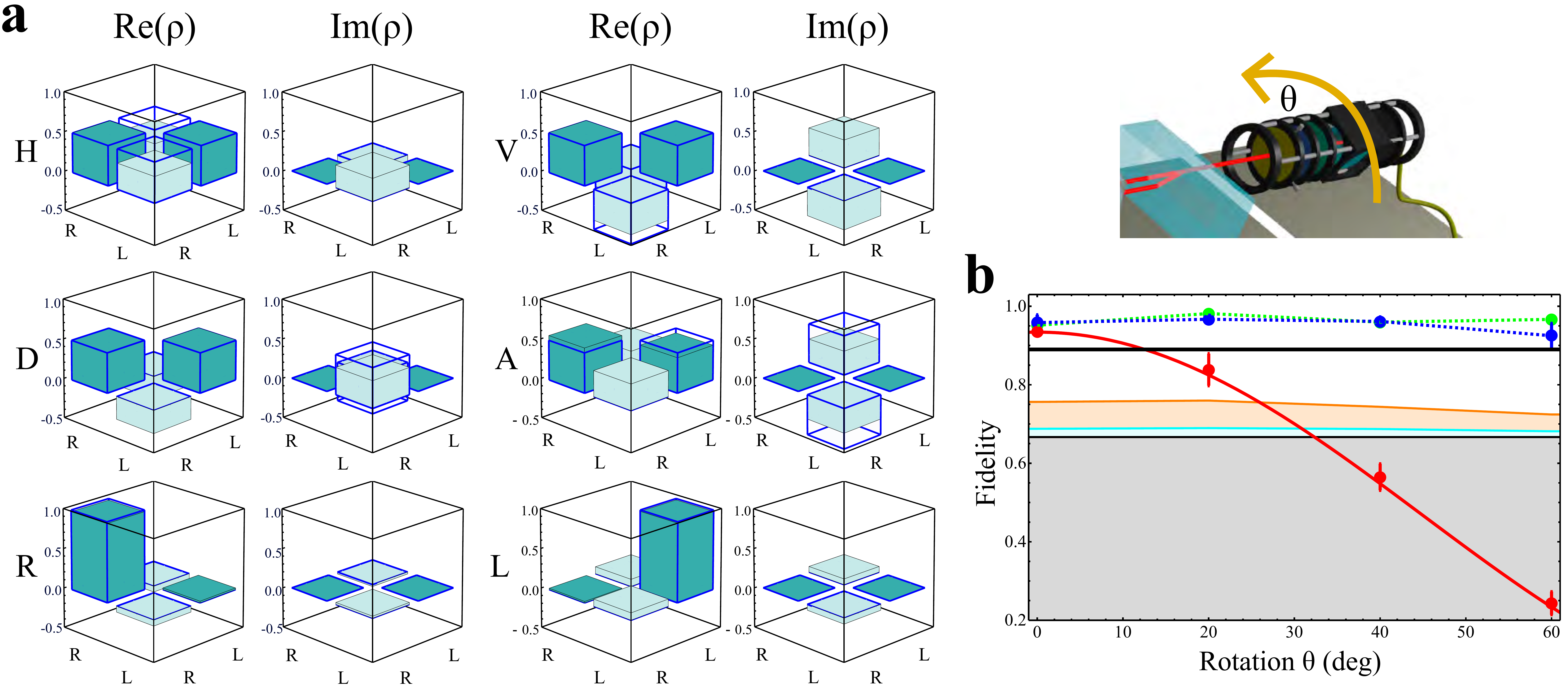}
\end{center}
\caption{\textbf{Conservation of rotational invariance by multiple-degree-of-freedom storage of vector beams.} (a) Tomography of polarization-encoded states after 1 $\mu$s storage for a rotation angle of the detection apparatus fixed at $\theta=20^{\circ}$.  No $q$-plates are used here. $\vert L\rangle$ and $\vert R\rangle$ stand for left and right circular polarizations while $\vert H\rangle$, $\vert V\rangle$, $\vert D \rangle= (1/\sqrt{2})(\vert H\rangle + \vert V\rangle) $ and $\vert A \rangle= (1/\sqrt{2})(\vert H\rangle - \vert V\rangle) $ denote linear polarizations.  The initial states are represented by empty columns with blue edges. For states encoded in linear polarizations, strong errors appear as expected. (b) Fidelity of the retrieved states for different encodings as a function of the rotation angle $\theta$. No background correction has been applied. The polarization-encoded states have different behaviors depending on whether they exhibit circular (green points, averaged fidelity over the two circular states) or linear polarizations (red points, averaged fidelity over the 4 linear states). The  circular states are indeed invariant under rotation while the linear ones cannot be decoded with high fidelity starting from $\theta=20 ^{\circ}$. The red solid line gives the expected theoretical Malus law $F_{\lbrace\theta=0\rbrace}\cos^{2}(\theta)$. In contrast, as given by the blue points, the fidelity of the vector encoded states, averaged here over the six input states belonging to the hybrid Poincar\'e sphere, remains above the Shor-Preskill $F_{T}=0.89$ threshold (black line) in the range $\theta=0^{\circ}-60^{\circ}$. This feature enables misalignement-immune quantum communication protocols. The classical fidelity thresholds are calculated taking into account the $\overline{n}=0.5$ mean-photon number and the finite retrieval efficiency (see Methods). The corresponding areas are displayed with the same color-code as in Fig. \ref{tom}. The small decrease in the fidelity value at $\theta=60^{\circ}$ can be mainly attributed to a technical misalignment due to this large rotation. Error bars indicate the standard deviations of fidelities for the considered sets of states.} \label{fid}
\end{figure*}

Having verified the capability of faithfully storing vector beams, we next focus on its use for a quantum information scenario. Recently, vector states as generated here have indeed been used for the realization of alignement-free quantum communication protocols \cite{DAmbrosio2012,vallone2014free}. In this case both the sender and the receiver write and analyze the qubit in the polarization space without sharing a reference frame since they use q-plates to switch between polarization and vector space spanned by the logical vectors $\ket{0}$ and $\ket{1}$. As these logical states are invariant under arbitrary rotation, the fidelity of these protocols is insensitive to rotations between the two reference systems used by the two parties. A possible more involved scenario could be the following: one of the two parties, Alice, prepares the state and sends it to the second party, Bob, who stores the state for a certain time and later on retrieves it to establish a secret key. Whatever the basis misalignment, the fidelity of the retrieved state must be greater than the $F_{T}=0.89$ value, which corresponds to the Shor-Preskill security-proof threshold \cite{Shor00} for the BB84 quantum key distribution (QKD) protocol. This feature can be experimentally tested by placing all the components of the detection apparatus (i.e. the second q-plate and polarization analysis elements) on a single support free to rotate around the propagation axis of the retrieved beam, as shown in Fig. \ref{fid}. We prepared and characterized state encoding either in polarization, as usually done in QKD protocols, or in hybrid state with various rotation angles $\theta$ of the detection cage. 

To first show the effect of misalignment on usual polarization qubits, we removed the two q-plates and stored such qubits. Fig. \ref{fid}a provides the reconstructed density matrices for the retrieved states with a rotation angle $\theta$ of the detection apparatus fixed at 20$^{\circ}$. As it can be seen for linear polarizations, large errors occur. Figure \ref{fid}b gives the achieved fidelity as a function of the rotation angle on a large range. The close-to-unity fidelity achieved at $\theta=0^{\circ}$ shows the faithful storage of polarization qubits in our memory device. As expected, when the angle increases, the fidelity decays for the linearly-polarized states following a Malus law while it stays close to its maximal value for circularly-polarized states.

Similar experiments were then performed for qubits encoded in the hybrid logical states, as superimposed in Fig. \ref{fid}b. In strong contrast to the polarization encoded states, all hybrid states after storage and decoding exhibit fidelities above the  $F_{T}$ threshold independently of the rotation angle. This result demonstrates the preservation of rotational invariance by the multiplexed storage and thereby the suitability of the realized memory for alignment-free quantum information protocols, with applications to quantum key distribution and long-distance repeaters where multimode quantum memories are a requisite building block.

We have demonstrated a quantum memory enabling the storage and retrieval of vector vortex beams, i.e hybrid polarization-OAM states, as a result of the spatially multimode nature of the ensemble-based implementation and of an additional dual-rail polarization multiplexing. This combination offers a multiple-degree-of-freedom register for light, at the single-photon level, with applications to quantum networks. Due to the spatial extent of the ensemble and the scaling of LG beam size in $\sqrt{d}$, where $d$ is the number of quanta of angular momentum, the current setup can be used for OAM space up to $d\sim50$. As an example, we have shown the conservation of rotational invariance for qubits encoded for misalignment-immune quantum communications. Our work thereby provides a novel capability for harnessing and further exploiting structured complex vector fields. Besides quantum information, further combinations of the peculiar properties of vector beams and light-matter interfacing protocols as demonstrated here should also lead to novel applications, such as, among others, possible ultra-sensitive magnetometers by using photonics gear based on higher topological charges \cite{DAmbrosio2013}.

\appendix
\section{Ensemble-based memory implementation}  
Memory experiments were driven at a repetition rate of 66 Hz, each cycle including a stage dedicated to MOT preparation and a period for memory operations (Fig. \ref{setup}c). The MOT preparation started with 11.5 ms of MOT loading followed by further cooling by optical molasses during 650 $\mu$s while the MOT magnetic field gradient was switched off. The optical depth decays then in a typical time constant of 2 ms and the memory efficiency stays thus almost constant over the memory period. Memory sequences were repeated 50 or 100 times during the memory operation part of the cycle (depending on the storage time), for a total number of acquisition of $150000$ for each projection. Temporal shaping of the pulses to be stored was obtained by applying a radio-frequency exponentially rising voltage on an acousto-optic modulator and photons were finally detected by a single avalanche photodiode (SPCM-AQR-14FC). In order to avoid inhomogeneous broadening, three pairs of coils were used to compensate any residual magnetic fields, down to 5 mG. The retrieval efficiency decays with storage time due to the decoherence of the collective atomic spin. Motional dephasing was the principal decoherence here. Due to the  $3 ^{\circ}$ angle between the signal and the control beams, the expected coherence time is around 7 $\mu$s, which is consistent with the experimental measurement.

\section{Assessing the quantum character of the memory} 
In order to assess the quantum character of the demonstrated memory, the measured fidelities have to be compared with the maximum fidelities achievable in a classical memory protocol, known for instance as the intercept-resend attack in quantum cryptography scenario. In the case of a N-photon state the maximal classical fidelity is given by $\alpha=(N+1)/(N+2)$, which leads to the well-known 2/3 limit for a single-photon state.
In the case of a coherent beam, as used here, the N-photon value $\alpha$ has to be averaged by taking into account the photon-number Poissonian distribution and the achievable fidelity can then be written as:
\begin{equation}
\sum_{N\geq 1}\frac{(N+1)}{(N+2)}\frac{P(\overline{n},N)}{1-P(\overline{n},0)}
\end{equation}
where $P(\overline{n},N)=e^{-\overline{n}} \overline{n}^{N}/N! $. The non-unity retrieval efficiency has also to be taken into account. A classical memory in an intercept-resend strategy could indeed simulate non-unity efficiencies to increase the achievable fidelity by giving an output only when the entering photon-number is above a certain threshold and inducing losses otherwise. Explicit expression for given mean-photon number and efficiency of the process are detailed in Refs. \cite{Specht11,Gundogan12}. The maximal classical fidelities have been reported in Fig. \ref{tom}b and \ref{fid}b 
where the blue and orange solid lines, which take into account  the Poissonian statistics and non-unity efficiency respectively, are extrapolated by calculating the best classical fidelity with the measured values of mean-photon number and
retrieval efficiency for every data set.
 
\section*{Acknowledgements}
\noindent The authors thank A. Nicolas, D. Maxein, E. Giacobino, L. Giner and L. Veissier for their contributions in the early stage of the experiment. The authors also acknowledge interesting discussions within the CAPES-COFECUB project Ph 740-12. This work was supported by the ERA-Net CHIST-ERA (QScale) and by the European Research Council (ERC Starting Grant HybridNet, grant agreement no. 307450  and ERC Starting Grant 3D-QUEST, grant agreement no. 307783, http://www.3dquest.eu.). J.Laurat is a member of the Institut Universitaire de France.\\

\end{document}